\documentclass[10pt,letterpaper]{article}
\usepackage{tb}
\usepackage{color}
\usepackage{amsmath}
\usepackage{cite}
\usepackage[]{graphicx} 
\usepackage{epstopdf}
\usepackage{ amssymb }
\usepackage{hyperref}

\begin{document}

\title{Phase-locking in cascaded stimulated Brillouin scattering}

\author{Thomas F. S. B\"{u}ttner$^{1,*}$, Christopher G. Poulton$^{2}$, M. J. Steel$^{3}$, \\Darren D. Hudson$^{1}$, Benjamin J. Eggleton$^{1}$}

\address{$^1$Centre for Ultrahigh bandwidth Devices for Optical Systems (CUDOS), Institute of Photonics and Optical Science (IPOS), School of Physics, University of Sydney, NSW, 2006, Australia\\
$^2$CUDOS, School of Mathematical Sciences, University of Technology, Sydney, NSW, 2007, Australia\\
$^2$CUDOS, MQ Photonics Research Centre, Department of Physics and Astronomy, Macquarie University, New South Wales 2109, Australia}

\email{$^*$thomasb@physics.usyd.edu.au} 



\begin{abstract}
Cascaded stimulated Brillouin scattering (SBS) is a complex nonlinear optical process that results in the generation of several optical waves that are frequency shifted by an acoustic resonance frequency. Four-wave mixing (FWM) between these Brillouin shifted optical waves can create an equally spaced optical frequency comb with a stable spectral phase, i.e. a Brillouin frequency comb 
(BFC). Here, we investigate phase-locking of the spectral components of BFCs, considering FWM interactions arising from the Kerr-nonlinearity as well as from coupling by the acoustic field. Deriving for the first time the coupled-mode equations that include all relevant nonlinear interactions, we examine the contribution of the various nonlinear processes to phase-locking, and show that different regimes can be obtained that depend on the length scale on which the field amplitudes vary.
\end{abstract}

\vspace{0.5cm} 




\section{Introduction}

When a narrow linewidth optical pump wave reaches a certain power threshold in a dielectric material, it can generate a counter-propagating optical Stokes wave together with a strong co-propagating acoustic wave through the process of backward stimulated Brillouin scattering (BSBS). The generated Stokes wave is Doppler red-shifted with respect to the pump wave by the acoustic resonance frequency $\Omega_B/2\pi$. The acoustic resonance frequency of BSBS depends mainly on material properties and on the optical wavelength and is on the order of 10 GHz for optical wavelengths of 1550 nm in glasses \cite{Boyd2008,Agrawal2001,Eggleton2013}. 
BSBS can be cascaded, in the sense that a Stokes wave acts as pump waves for higher order Stokes waves. This can be exploited to achieve comb-like optical spectra that contain several spectral components spaced by the acoustic resonance frequency $\Omega_B/2\pi$. Cascaded BSBS has been observed in several resonator configurations that include hybrid erbium-Brillouin fiber ring lasers \cite{Cowle1996a}, Fabry-Perot (FP) resonators \cite{Braje2009,Pant2011a,Buttner2014,Buettner2014b} and whispering gallery mode resonators \cite{Tomes2009,Grudinin2009,Li2012a}. In recent demonstrations of cascaded BSBS in FP resonators, stable phase-locking between the optical waves has been observed \cite{Braje2009,Buttner2014,Buettner2014b}, i.e. the generation of Brillouin frequency combs (BFCs).

Early theoretical studies of cascaded BSBS in resonators did not consider the Kerr-nonlinearity and the finite linewidth of the acoustic resonance \cite{Korolev1973,Lugovoi1983}. These studies showed that, in certain configurations,  optical waves can be phase-locked by phase-sensitive FWM interactions that are mediated by the acoustic field. Such interactions have been referred to as Brillouin enhanced FWM (BEFWM) \cite{Scott1989}. Several more recent theoretical studies on cascaded BSBS \cite{Lecoeuche1996,Stepanov1999,Ogusu2002a,Ogusu2002b,Ogusu2003} did not include phase-sensitive FWM interactions or used coupled power equations and are therefore not suitable to explain the phase-locking observed in experiments. Recently, we numerically showed that FWM interactions arising from the Kerr-nonlinearity (KFWM) between co-propagating waves can lead to phase-locking of the optical wave generated by cascaded BSBS in a 38 cm long chalcogenide fiber resonator~\cite{Buttner2014}. In this numerical study, the acoustic field was represented as a superposition of several acoustic waves following the approach of reference~\cite{Ogusu2002a} and BEFWM interactions were not included since the fiber under investigation was much longer than the coherence length of BEFWM ($\approx 7$ mm \cite{Buttner2014,Buettner2014b}). 

Here we provide for the first time, a detailed derivation of the dynamic coupled-mode equations for cascaded BSBS in fibers, which includes all phase-sensitive KFWM and BEFWM interactions. Such equations are necessary to correctly describe phase-locked cascaded BSBS for a wide range of resonator lengths. The full equations are derived by describing the acoustic field with a single acoustic wave as well as with a superposition of several acoustic waves. The validity of both approaches is demonstrated,
thereby resolving questions that have recently arisen over the appropriate formalism \cite{Broderick2015}. 
We use our equations to examine the relative importance of the different FWM interactions with regard to phase-locking and numerically simulate cascaded BSBS in low-finesse FP resonators of different lengths. We find that depending on the length scale on which the field amplitudes vary, different FWM interactions are relevant for phase-locking of the BFCs. 

\section{Coupled-mode equations for cascaded BSBS}\label{Sec:cmes_all_waves}

In this section, the coupled-mode equations for cascaded BSBS are derived for a weakly guiding, single-optical-mode dielectric waveguide consisting of an isotropic material. The waveguide is oriented along the $z$-direction and has transverse dimensions that are large enough so that boundary effects can be neglected. Furthermore, we restrict our analysis to a narrow optical bandwidth and assume that only one polarization mode is excited and that the state of polarization is maintained. 

In optical fibers, the acoustic pressure waves participating in BSBS are well represented by the density variation $\rho(\bold{r},t)$ from the mean density of the medium $\rho_0$. The wave equation for $\rho(\bold{r},t)$ that governs the propagation of the acoustic wave can be written in the form \cite{Boyd2008,Kobyakov2009}
\begin{align}\label{eq:theory_acoustic6}
\frac{\partial^2 \rho}{\partial t^2}-v_a^2\nabla^2\rho-\Gamma \nabla^2 \frac{\partial}{\partial t}\rho =-\frac{1}{2} \varepsilon_0\gamma_e \nabla^2 E^2+\tilde{f},
\end{align}
where $v_a$ is the velocity of an acoustic pressure wave, $\Gamma$ is a damping parameter,  $\varepsilon_0 $ is the vacuum permittivity and $\gamma_e$ is the electrostrictive constant. $\tilde{f}$ represents a Langevin noise source associated with the acoustic dissipation that is necessary to include thermal excitation of acoustic waves \cite{Boyd1990}.

The optical waves are represented in a scalar picture by the electric field $E(\bold{r},t)$. The propagation of the electric field is governed by the wave equation \cite{Agrawal2001,Kobyakov2009}
\begin{align}\label{eq:theory_el_wave}
{\nabla ^2}E - \frac{{{\varepsilon_\text{L}}}}{{{c^2}}}\frac{{{\partial ^2}E}}{{\partial {t^2}}} = \frac{1}{{{\varepsilon _0}{c^2}}}\frac{{{\partial ^2}{P_{\text{NL} }}}}{{\partial {t^2}}},
\end{align}
where $c$ is the vacuum speed of light. The parameter ${\varepsilon _L}$ is the linear part of the relative permittivity that can be written in the form \cite{Agrawal2001,Kobyakov2009}
\begin{align}\label{eq:epsilon_l}
{\varepsilon _L} = {n^2} + in\alpha c/\omega_0,
\end{align}
where $n$ is the refractive index, $\alpha$ is the attenuation coefficient of the optical waves and $\omega_0$ is the angular frequency of the optical wave. Here, the parameters $n$ and $\alpha$ are assumed to be wavelength independent due to the narrow optical bandwidth that is considered. $P_{\text{NL}}$ in equation \ref{eq:theory_el_wave} represents the nonlinear part of the polarization density. In our analysis we include the contribution to $P_{\text{NL}}$ arising from the optical Kerr-effect (described by the third order susceptibility $\chi ^{(3)}$) and from the variation of the material density $\rho(\bold{r},t)$ resulting from the presence of acoustic waves:
\begin{align}\label{eq:P_NL}
   P_{\text{NL}}=\varepsilon_0 \varepsilon_\text{NL}E= \varepsilon_0\left(\chi ^{(3)}{E^2} + \frac{{{\gamma _E}}}{{{\rho _0}}}\rho \right) E,
\end{align}
where $\varepsilon_\text{NL}$ is the nonlinear part of the relative permittivity. Assuming that the nonlinear polarization can be treated as a small perturbation to the total polarization and that the interaction takes place over a narrow optical bandwidth, we use the following approximation \cite{Kobyakov2009}
\begin{align}\label{eq:P_NL_approx}
    \frac{\partial^2}{\partial t^2} P_{\text{NL}}\approx \varepsilon_0 \varepsilon_\text{NL} \frac{\partial^2}{\partial t^2} E.
\end{align}

\begin{figure}[!t] \centering
 \includegraphics[width=0.5\linewidth]{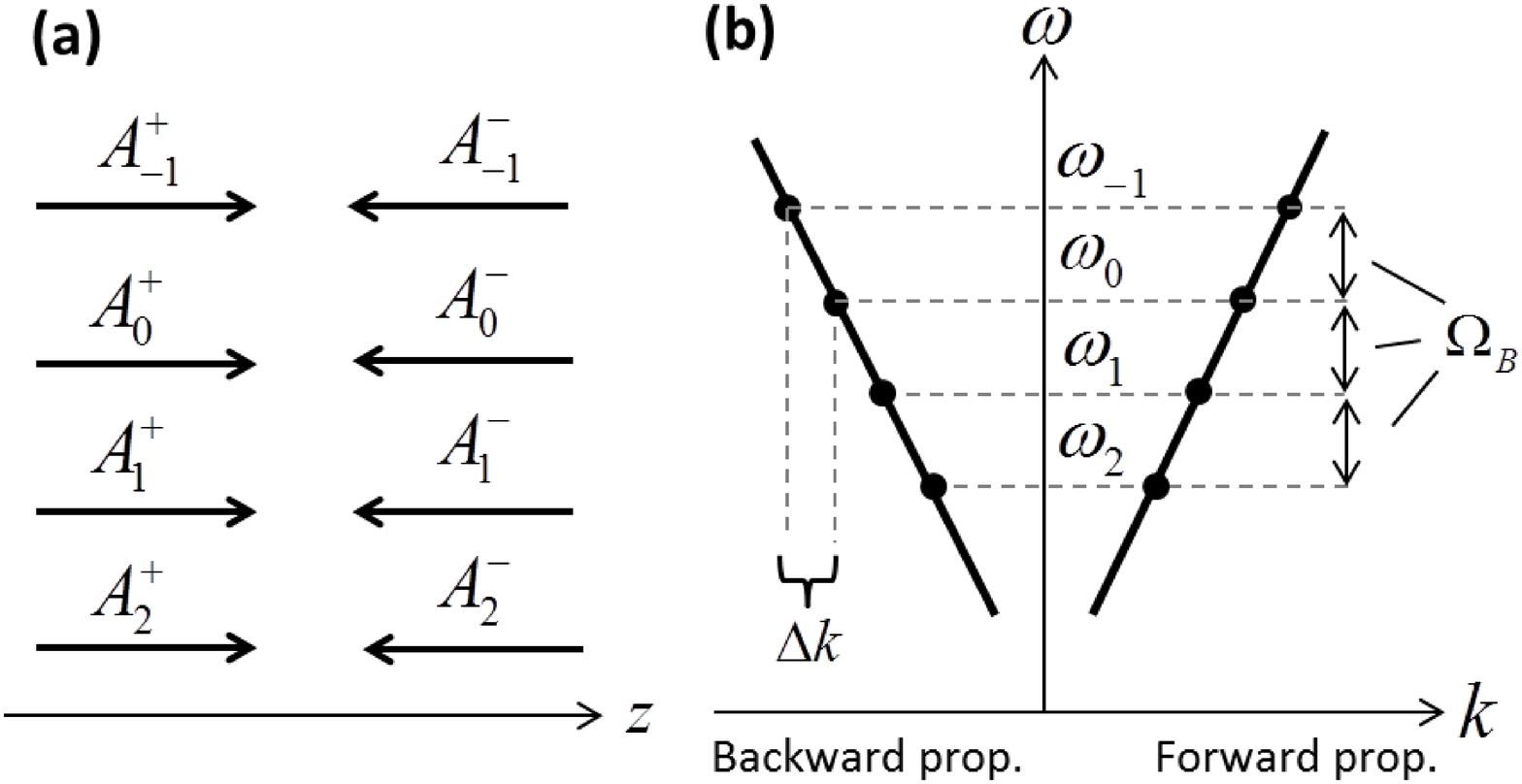}
  \caption{(a) Schematic diagram showing the propagation direction of the optical waves. (b) Energy-momentum diagram illustrating the definition of the frequencies and the momenta of the optical waves (black dots) and the propagation constant difference $\Delta k$.}
    \label{fig:concept}
\end{figure}

We describe the electric field as superposition of forward and backward propagating waves with propagation constants $\pm k_j$, angular frequencies $\omega_j$ and amplitudes $A_j^ \pm (z,t)$:
\begin{align}\label{eq:electric_theory_paper}
 E(\bold{r},t) = 
 \frac{1}{2}F(x,y)C_E\sum\limits_j {\left( {A_j^ + (z,t){e^{i{k_j}z}} + A_j^ - (z,t){e^{ - i{k_j}z}}} \right)\;} {e^{ - i{\omega _j}t}} + c.c.
\end{align}
The index $j=0$ denotes the pump wave, indexes $j\geq1$ Stokes waves and indexes $j\leq-1$ anti-Stokes waves. The carrier frequencies of the pump wave and the order first Stokes wave are defined such that their frequency spacing ${\omega _0} - {\omega _1}$  corresponds exactly to the acoustic resonance frequency $\Omega_B$ of a pressure wave that travels with velocity ${v_a}$ along the $z$-axis
\begin{align}\label{eq:theory_def_omegaB}
{\omega _0} - {\omega _1} = {\Omega _B} = {K_0}{v_a} ,
\end{align}
where $K_0={k_0} + {k_1}$ is the corresponding propagation constant of the acoustic wave. The angular frequencies of all higher order Stokes and anti-Stokes waves are defined as
\begin{align}\label{eq:fiber_def_stokes_frequencies}
{\omega _j} = {\omega _0} - j \times {\Omega _B}, 
\end{align}
i.e.~exactly equally spaced by the acoustic frequency $\Omega _B$ for the interaction between the pump and the first order Stokes wave (see Fig.~\ref{fig:concept}). 

The parameter $C_E=(\varepsilon_0 c n \left< F^2 \right>_{x,y}/2)^{-1/2}$ in equation~\ref{eq:electric_theory_paper} scales the optical amplitudes such that $|A_j^\pm|^2$ corresponds to the optical power carried by each mode. $F(x,y)$ represents the transverse mode profile for guided optical modes with frequency close to $\omega_0$ and fulfills the modal equation
\begin{align}\label{eq:bg_modal_optical}
\left[ {\partial _x^2 + \partial _y^2 + \frac{n^2(x,y)\omega_0^2}{c^2}- {k_0}^2} \right]F(x,y) = 0.
\end{align} 
Note that in \eqref{eq:electric_theory_paper} we neglect the frequency dependence of the field profile $F(x,y)$ because of the small frequency range involved in Brillouin interactions.
The angular brackets $\left< ..\right>_{x,y}$ denote averaging over the transverse plane, using the integral $\iint..\,\mathrm{d}x\mathrm{d}y$.

\subsection{Derivation with a single acoustic wave}

In this section, we derive the coupled-mode equations for cascaded BSBS by expressing the density variation $\rho(\bold{r},t)$ as superposition of only one forward and one backward propagating acoustic wave with propagation constants $\pm K_0$ and angular frequency $\Omega_B$:
\begin{align}\label{eq:acoustic_theory_paper}
 \rho (\bold{r},t) = \frac{1}{2}\xi (x,y)\left[ {{Q^ + }(z,t){e^{iK_0z}} + {Q^ - }(z,t){e^{-i  K_0z }}} \right]e^{-i\Omega_B t} + c.c.,
\end{align}
where ${Q^+ }(z,t)$ and ${Q^ - }(z,t)$ are the amplitudes of forward and backward propagating acoustic waves, respectively. In general many acoustic modes exist. However, here we only consider the interaction with the fundamental longitudinal acoustic mode that dominates in the SBS interaction  and has mode profile $\xi_0 (x,y)$. $\xi_0 (x,y)$ fulfills the modal equation
\begin{align}\label{eq:bg_modal_acoustic}
\left[ {\partial _x^2 + \partial _y^2 + \frac{\Omega_B^2}{v_a(x,y)^2} - {K_0^2}} \right]\xi (x,y) = 0.
\end{align}
In order to derive the coupled-mode equations for the optical waves, we start by inserting equations~\ref{eq:electric_theory_paper} and~\ref{eq:acoustic_theory_paper} into equation~\ref{eq:theory_el_wave} and substituting equations~\ref{eq:epsilon_l} and \ref{eq:P_NL}. 
We then use the approximation shown in equation~\ref{eq:P_NL_approx} and apply the slowly varying envelope approximation by omitting second derivatives of the amplitudes \cite{Agrawal2001,Kobyakov2009}.
 We separate the parts with exponential factors ${e^{i({k_j}z - {\omega _j}t)}}$ and ${e^{i( - {k_j}z - {\omega _j}t)}}$ into the following equations 
\begin{align}\label{eq:theory_cme_optical_1acoustic}
  &  \pm {\partial _z}A_j^ \pm  + \frac{n}{c}{\partial _t}A_j^ \pm + \frac{\alpha }{2}A_j^ \pm  =   \cr 
  &  =   ig_2 \left( {{Q^ \pm }A_{j + 1}^ \mp {e^{ \pm i(2j\Delta k)z}} + {Q^ \mp }^*A_{j - 1}^ \mp {e^{ \pm i2(j - 1)\Delta kz}}} \right)  \cr
  &  + i\gamma \sum\limits_{l,m,p} {{\delta _{p,{\kern 1pt} l + m - j}}\left( {A_l^ \pm A_m^ \pm A{{_p^ \pm }^*} + 2A_l^ \pm A_m^ \mp A{{_p^ \mp }^*}{e^{ \pm i2(j - l)\Delta kz}}} \right)}
\end{align}
with
\begin{align}
\gamma & = \frac{3 \text{Re}(\chi^{(3)})\omega_0 {\left\langle {{F^4} } \right\rangle_{x,y} }}{8nc{\left\langle {{F^2} } \right\rangle_{x,y} }}C_E,\\
g_2&=\frac{{\omega_0 {\gamma _e}}}{{4nc{\rho _0}}}\frac{{\left\langle {{F^2}\xi } \right\rangle_{x,y} }}{{\left\langle {{F^2}} \right\rangle_{x,y} }}.
\end{align}
To derive equation~\ref{eq:theory_cme_optical_1acoustic}, we substituted the modal equation equation~\ref{eq:bg_modal_optical}. In order to separate the transverse dependency, both sides of the equation were multiplied by $F(x,y)$, integrated over the transverse plane and divided by $\left<F^2\right>_{x,y}$. Furthermore, we used the approximations $\omega_j  \approx \omega_0 $ and ${k_j} \approx {\omega _j}n/c$ in prefactors since we only consider a small frequency range and a weakly guiding waveguide.

The source term on the right hand side of equation~\ref{eq:theory_cme_optical_1acoustic}, containing the parameter ${g_2}$, describes the coupling of the optical waves $A_j^ \pm$   to the counter propagating waves $A_{j + 1}^ \mp $ (Stokes processes)  and $A_{j - 1}^ \mp $ (anti-Stokes processes) via the acoustic waves ${Q^ \pm }$  and ${Q^ \mp }$, respectively. Here, the $z$  and $j$  dependent phase rotations result from the propagation constant mismatches 
\begin{align}
2j\Delta k = K_0 - k_j-k_{j+1},
\end{align}
where $\Delta k$ represents the propagation constant difference between two adjacent spectral components
\begin{align}
\Delta k = {k_j} - {k_{j + 1}} = {\Omega _B}n/c.
\end{align}
In equation~\ref{eq:theory_cme_optical_1acoustic}, the term containing $\gamma $ represents the interactions via the Kerr-nonlinearity. The sum includes all combinations of the indexes $l,m$ and $p$ and the Kronecker delta ${\delta _{p,{\kern 1pt} l + m - j}}$   ensures conservation of energy. The first term of the sum includes KFWM of co-propagating waves, which has no propagation constant mismatch (under zero dispersion assumption). The second term of the sum contains KFWM of counter-propagating waves, which has a propagation constant mismatch of $2(j - l)\Delta k$.

We proceed similarly for deriving the equations for the amplitudes of the acoustic waves and insert equations~\ref{eq:electric_theory_paper} and~\ref{eq:acoustic_theory_paper} into equation~\ref{eq:theory_acoustic6}. We then use the slowly varying envelope approximation and separate the terms with the exponential factors ${e^{i(  {K_0}z - {\Omega_B}t)}}$ and ${e^{i( - {K_0}z - {\Omega_B}t)}}$ to obtain 
\begin{align}\label{eq:theory_cme_acoustic_1acoustic}
   {\partial _t}{Q^ \pm } \pm {v_a}{\partial _z}{Q^ \pm }+ \frac{\Gamma_B}{2}{Q^ \pm }= {ig_1\sum\limits_j {A_j^ \pm A{{_{j + 1}^ \mp }^*}{e^{ \mp i2j\Delta kz}}}   } +f^\pm,
\end{align}
with
\begin{align}
g_1&=\frac{{\Omega_B {\gamma _e}}}{{2v_a^2cn}}\frac{{\left\langle {\xi {F^2}} \right\rangle_{x,y} }}{{\left\langle {{F^2}} \right\rangle_{x,y} \left\langle {{\xi ^2}} \right\rangle_{x,y} }}.
\end{align}
In equation~\ref{eq:theory_cme_acoustic_1acoustic}, we have substituted the decay constant of the acoustic wave $\Gamma_B = {(\Gamma {K_0^2})}$, which is related to the phonon lifetime $\tau_p$ with $\Gamma_B=\tau_p^{-1}$. In deriving equation~\ref{eq:theory_cme_acoustic_1acoustic}, we used the approximation $v_a(k_j+k_{j+1})\approx \Omega_B$ for prefactors. Several terms that contain first derivatives of the amplitudes $Q^ \pm$ have been neglected based on the approximation $\Gamma_B /\Omega_B \approx 0$, which represents the ratio between the Brillouin gain bandwidth and the Brillouin frequency shift \cite{Kobyakov2009}. We also substituted the modal equation for the acoustic wave (equation~\ref{eq:bg_modal_acoustic}) and separated the transverse dependency by multiplying both sides of the equation with $\xi_0 (x,y)$, integrating over the transverse plane and dividing by $\left<\xi^2_0\right>_{x,y}$.

The first term on the right hand side of equation~\ref{eq:theory_cme_acoustic_1acoustic} shows that the acoustic waves ${Q^ \pm }$ are driven by all pairs of optical waves $A_j^ \pm $ and $A_{j + 1}^ \mp $. This term also contains the propagation constant mismatch $2j\Delta k$ since the optical waves drive the acoustic field with different propagation constants $k_j+k_{j+1}$.

The noise terms $f^\pm$ equation~\ref{eq:theory_cme_acoustic_1acoustic} are related to the original noise term $\tilde{f}$ by
\begin{align}
\tilde{f} = -i \Omega_B \xi_0(x,y) \left( {{f^ + }(z,t){e^{iK_0z }} + {f^ - }(z,t){e^{-iK_0z }}} \right)e^{- i\Omega_B t}+ c.c.
\end{align}
The noise sources $f^+$ and $f^-$ are uncorrelated Gaussian random variables with zero mean $\left<f^\pm\right>=0$ and are $\delta$ correlated in the sense
\begin{align}\label{eq:theory_ch_noise_strenght}
\left<f^\pm(z,t) f^\pm(z',t')\right>=\psi \delta(z-z')\delta(t-t'),
\end{align} 
where the parameter $\psi$ describes the strength of the thermal fluctuations. An expression for $\psi$ is given by equation 18 in reference \cite{Boyd1990} which is based on the equipartition theorem.   

Note that in equations~\ref{eq:theory_cme_optical_1acoustic} and \ref{eq:theory_cme_acoustic_1acoustic}, the phase of the terms containing the propagation constant mismatch $2j\Delta k$ rotates by $j \times 2\pi $  over the ``Brillouin-length'' \cite{Lugovoi1983}
\begin{align}\label{eq:bg_lb}
L_B = \frac{\pi c}{\Omega_Bn}=\frac{\pi}{\Delta k},
\end{align}
which is about 1 cm for the parameters of standard silica fiber at 1550 nm \cite{Agrawal2001}.

Equations~\ref{eq:theory_cme_optical_1acoustic} and~\ref{eq:theory_cme_acoustic_1acoustic} represent the coupled-mode equations for cascaded BSBS derived using only one forward and one backward propagating acoustic wave.

\subsection{Derivation with several acoustic waves}

\begin{figure}[!b] \centering
 \includegraphics[width=0.6\linewidth]{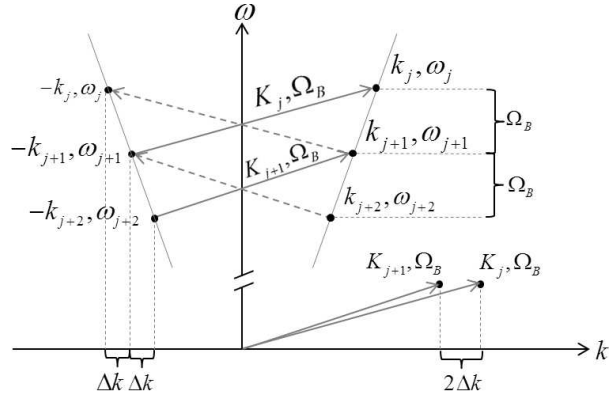}
  \caption{Energy-momentum diagram that schematically illustrates the propagation constants and angular frequencies of the optical and acoustic waves and the propagation constant mismatch $2j\Delta k$.}
    \label{fig:theory_dispersion}
\end{figure}

It is common for modeling cascaded backward BSBS to describe the acoustic field as superposition of several acoustic waves \cite{Ogusu2002a,Buttner2014}. This approach can be insightful and useful for numerical modeling of waveguides with lengths $L\gg L_B$. To derive the corresponding coupled-mode equations, we express the acoustic field $\rho (\bold{r},t)$ as the superposition of several acoustic waves with different propagation constants (corresponding to the propagation constants of the driving fields) 
\begin{align}\label{eq:theory_propagation_constant_acoustic_waves}
 {K_l} = {K_0} - 2l\Delta k = {k_l} + {k_{l + 1}}
 \end{align}
 but with the same temporal frequency $\Omega_B={K_0}{v_a}$ (see Fig.~\ref{fig:theory_dispersion}). The density variation then reads
\begin{align}\label{eq:BFC_fiber_multiple_acoustic_def}
\rho (\bold{r},t)=  \frac{1}{2}\xi_0  (x,y){e^{ - i{\Omega _B}t}} \sum\limits_l {\left[ {Q_l^ + (z,t){e^{i{K_l}z}} + Q_l^ - (z,t){e^{ - i{K_l}z}}} \right]} + c.c.
\end{align}

To derive the coupled-mode equations, we insert equations~\ref{eq:electric_theory_paper} and~\ref{eq:BFC_fiber_multiple_acoustic_def} into equations~\ref{eq:theory_el_wave} and~\ref{eq:theory_acoustic6} and proceed as in the previous section to derive
\begin{align}\begin{split}\label{eq:bfc_fiber_aj_with_qls}
  &  \pm {\partial _z}A_j^ \pm  + \frac{n}{c}{\partial _t}A_j^ \pm + \frac{\alpha }{2}A_j^ \pm    =   \\
  &  =   ig_2\sum\limits_l {\left( {Q_l^ \pm A_{j + 1}^ \mp  + Q{{_{l - 1}^ \mp }^*}A_{j - 1}^ \mp } \right){e^{ \pm i2(j - l)\Delta kz}}}   \\ &  + i\gamma \sum\limits_{l,m,p} {{\delta _{l + m - p - j}}\left( {A_l^ \pm A_m^ \pm A{{_p^ \pm }^*} + 2A_l^ \pm A_m^ \mp A{{_p^ \mp }^*}{e^{ \pm i2(j - l)\Delta kz}}} \right)},  
  \end{split} 
\\
\begin{split}\label{eq:bfc_fiber_several_acoustic}
\sum\limits_l {\left[ {{\partial _t}Q_l^ \pm  \pm {v_a}{\partial _z}Q_l^ \pm } +\frac{\Gamma_B}{2}(1 - il\delta )Q_l^ \pm\right]}e^{-il \Delta k}   
  = ig_1 \sum\limits_l { {A_l^ \pm A{{_{l + 1}^ \mp }^*}   } e^{-il \Delta k}+f^{\pm}}. 
\end{split}
\end{align}
In equation~\ref{eq:bfc_fiber_several_acoustic}, only forward and backward propagating acoustic waves were separated. The detuning term containing $l\delta  = 4l\Delta k{v_a}/\Gamma_B $ now appears in the equation because the acoustic waves with propagation constants, as defined in equation~\ref{eq:theory_propagation_constant_acoustic_waves}, do not obey the acoustic dispersion relation $v_aK_l\neq \Omega_B$, except for  $l = 0$. 

In order to obtain separate equations, we define $Q_l^\pm$  to obey the following equation
\begin{align}\label{eq:bfc_fiber_ql}
{\partial _t}Q_l^ \pm  \pm {v_a}{\partial _z}Q_l^ \pm  +\frac{\Gamma_B}{2} (1 - il\delta )Q_l^ \pm =  {ig_1 A_l^ \pm A{{_{l + 1}^ \mp }^*}  } +f_l^\pm,
\end{align}
where $f_l^\pm=f^\pm e^{\pm il\Delta k }/N$ and $N$ is the number of acoustic waves. By comparing equations~\ref{eq:bfc_fiber_several_acoustic} and~\ref{eq:bfc_fiber_ql}, it can be seen that when $Q_l^ \pm (z,t)$ are defined as in  equation~\ref{eq:bfc_fiber_ql}, equation~\ref{eq:bfc_fiber_several_acoustic} is also fulfilled. However, it should be noted that the amplitudes $Q_l^ \pm (z,t)$, in general, are not uniquely defined by equation~\ref{eq:bfc_fiber_several_acoustic}. The different components of the acoustic field $Q_l^ \pm (z,t){e^{i(\pm {K_l}z - {\Omega _B}t)}}$ are not always orthogonal functions because they have the same frequency $\Omega _B$ and the spatial spectra of $Q_l^ \pm (z,t){e^{\pm i{K_l}z}}$ can also overlap if the amplitudes $Q_l^ \pm$ vary on a length scale $\sim L_B$. With equation~\ref{eq:bfc_fiber_ql}, we have defined the waves $Q_l^\pm$ as the parts of the acoustic field that are driven by the pairs of adjacent spectral components of the optical field $A_l^ \pm$ and  $A{_{l + 1}^ \mp }$.

\begin{figure}[!b]\centering
  \includegraphics[width=0.6\linewidth]{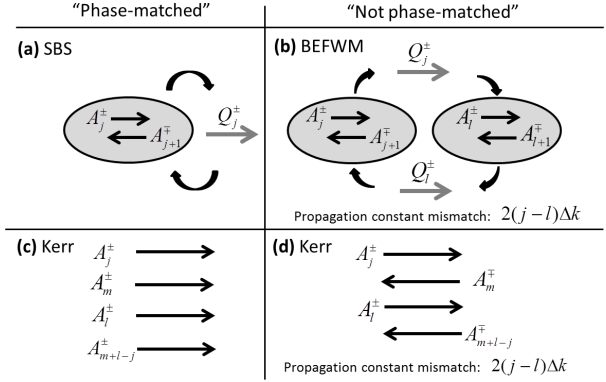}
  \caption{Illustration of the different categories of nonlinear interactions in cascaded BSBS in an FP resonator. (a) ``Common'' three-wave interaction of SBS  that is insensitive to the spectral phase. (b) - (d) show FWM interactions that are sensitive to the spectral phase: (b) BEFWM of counter-propagating optical waves,  which is mediated by the acoustic field, (c) KFWM of co-propagating waves and (d) KFWM of counter-propagating waves. Note that the FWM interactions of counter-propagating waves shown in (b) and (d) have a propagation constant mismatch of $2(j-l)\Delta k$.   }
    \label{fig:theory_fwm_illustration}
\end{figure}

The sum over $l$, containing the acoustic waves $Q_l^ \pm $ in equation~\ref{eq:bfc_fiber_aj_with_qls}, represents the coupling of the optical waves $A_j^ \pm $ to the counter-propagating waves $A_{j + 1}^ \mp $  and $A_{j - 1}^ \mp $, which is mediated by all acoustic waves $Q_l^ \pm $. This coupling is only phase-matched for $j = l$, which results from the phase-rotation terms containing  $j - l$. For $j = l$, there is no propagation constant mismatch as these terms then describe the coupling of two optical waves, e.g.~$A_j^ \pm $  and $A_{j + 1}^ \mp $,  by the acoustic wave $Q_j^ \pm $  that is also driven by the same pair of optical waves (Fig.~\ref{fig:theory_fwm_illustration} (a)), which is the usual three-wave interaction of BSBS. 
The other terms of the sum with  $j \ne l$, on the other hand, represent terms that we refer to as BEFWM interactions \cite{Narum2007,Scott1989}. One example of such an interaction is illustrated in Fig.~\ref{fig:theory_fwm_illustration} (b). 
The waves  $A_j^ \pm $  and $A_{j + 1}^ \mp $ are coupled by the acoustic wave $Q_l^ \pm $  with wavevector ${K_l} $,  which is driven by the waves $A_l^ \pm $   and $A_{l + 1}^ \mp $. At the same time, $A_l^ \pm $  and $A_{l + 1}^ \mp $ are coupled by the acoustic field $Q_j^ \pm $  with wavevector ${K_j} $, driven by the waves $A_j^ \pm $   and $A_{j + 1}^ \mp $.  This interaction can be efficient, e.g.~when the acoustic waves $Q_l^ \pm (z,t){e^{\pm i{K_l}z}}$ have significant strength at frequency ${k_j} + {k_{j + 1}}$  in  momentum-space, which is possible when $Q_l^ \pm (z,t)$  vary on a length scale $\sim{L_B}/|j - l|$. On the other hand, BEFWM interactions have no significant contribution if the spatial spectra of the acoustic waves $Q_l^ \pm (z,t){e^{\pm i{K_l}z}}$ and $Q_j^ \pm (z,t){e^{\pm i{K_j}z}}$ are well separated in momentum-space, which is the case when the amplitudes vary on a lengthscale $\gg L_B$. In this case, these terms can be omitted as previously been done \cite{Buttner2014,Ogusu2002a,Ogusu2003} (note that in this case the correct inclusion of the noise terms $f^\pm_l$ has to be reconsidered). For comparison, Figs.~\ref{fig:theory_fwm_illustration} (c) and (d) show the KFWM interactions of co-propagating waves and counter-propagating waves, respectively, as described in the discussion of equation~\ref{eq:theory_cme_optical_1acoustic}.

The peak Brillouin gain coefficient ${g_0}=4g_1g_2/\Gamma_B$ (units: m$^{-1}$W$^{-1}$) is often much larger than the nonlinear coefficient $\gamma $, e.g. in optical fibers. The phase-matched BSBS three-wave interaction (Fig.~\ref{fig:theory_fwm_illustration} (a)) is then expected to be the dominant gain process as it is phase-matched and has the larger gain constant. This interaction, however, cannot lead to phase-locking as it only couples pairs of optical waves and is independent of the spectral phase. The FWM interactions, on the other hand, couple more than two optical waves and are sensitive to the spectral phase. In the next section, we will numerically investigate the relative importance of the different FWM interactions with respect to phase-locking.

\section{Numerical analysis of cascaded BSBS in low-finesse Fabry-Perot resonators}\label{sec:fiber_numerical}

In this section, the dynamic coupled-mode equations, derived in the previous section, are used to study cascaded BSBS in low-finesse, FP fiber resonators of two different lengths. We qualitatively investigate the importance of the different FWM interactions for phase-locking the optical waves and its dependence on the resonator length. For simplicity, we only include forward and backward propagating pump, first and second order Stokes waves and the corresponding acoustic waves that couple the optical waves ($j=0,1,2$). 

The dynamic coupled-mode equations for cascaded BSBS were solved using the method of characteristics \cite{DeSterke1991} adapted to include additional acoustic waves \cite{Ogusu2002a,Winful2013a,Winful2013}. We used the coupled-mode equations~\ref{eq:bfc_fiber_aj_with_qls} and~\ref{eq:bfc_fiber_ql}, omitting the propagation of the acoustic wave (the term $\pm {v_a}{\partial _z}Q_l^ \pm $), which is a common approximation because the acoustic waves are highly damped \cite{Boyd2008,Kobyakov2009}. For a complete description of the system, the boundary conditions have to be defined. In the case of a FP resonator, the boundary conditions can be written as \cite{Ogusu2002a}
\begin{align}\begin{split}
  & A_0^ + (z = 0,t) = \sqrt R A_0^ - (0,t) + \sqrt {{P_{{\text{in}}}}},   \cr 
  & A_j^ + (z = 0,t) = \sqrt R A_j^ - (0,t){\text{      for $j$ = 1,2}},  \cr 
  & A_j^ - (z = L,t) = \sqrt R A_j^ + (L,t){e^{i\Delta \varphi_j }}{\text{  for  $j$ = 0,1,2}}. \cr
\end{split}\end{align}
The parameter ${P_{{\text{in}}}}$  is the pump power that is coupled into the resonator at  $z = 0$. $\Delta {\varphi _j} $ are the linear phase shifts of the waves due to transit through the cavity, which are given by
\begin{align}\label{eq:bg_transit_phase-shifts}
\Delta {\varphi _j} =2Lk_j= 2L(k_0-j \Delta k)=\Delta \varphi_0 - 2\pi j M.
\end{align}
The parameter $M$ can be described as
\begin{align}\label{eq:bg_def_M}
M = \frac{\nu _B}{\text{FSR}} = \frac{L}{L_B}.
\end{align}
 Here, we assume that the parameter $M$ is an integer, so that all waves effectively experience the same linear phase shift 
\begin{align}
\Phi=\Delta {\varphi _j}\text{mod}{\mkern 1mu} (2\pi )
\end{align}
after one roundtrip, where ``mod'' (modulo) denotes the remainder of $\Delta {\varphi _j}$ after division by $2\pi $.  Additionally, it is assumed that initially at $t=0$ all fields are zero inside the resonator. 

To identify phase-locking between the waves, we consider the phases $\varphi _j^ \pm (z,t)$ of the amplitudes 
\begin{align}
A_j^ \pm (z,t) = \sqrt {P_j^ \pm (z,t)}  e^{i\varphi _j^ \pm (z,t)},
\end{align}
where $P_j^ \pm (z,t)$ are real and represent the powers of the waves. 

The frequencies of the Stokes waves in the steady state do not necessarily coincide with the frequencies defined by equation $\ref{eq:fiber_def_stokes_frequencies}$ due to ``Stokes pulling'' (the Stokes frequencies lie in between the cavity resonances and the Brillouin gain peak \cite{Lecoeuche1996}). However, when the system reaches a phase-locked steady state, the individual phases can be expressed in the form  $\varphi _j^ \pm (z,t) = \tilde \varphi _j^ \pm (z) - \Delta {\omega _j}t$, where $\Delta \omega_j$ are frequency detunings that are $z$ and $t$  independent  \cite{Lecoeuche1996}. This means that the actual steady state Stokes and anti-Stokes frequencies are given by  ${\omega _j} + \Delta {\omega _j}$.

When the three waves reach a steady state with equally spaced frequencies, the phase-dispersion
\begin{align}
\vartheta ^ \pm (z,t) = \varphi _0^ \pm (z,t) + \varphi _2^ \pm (z,t) - 2\varphi _1^ \pm (z,t)
\end{align} 
is time independent, i.e.~$(d/dt)\vartheta _{}^ \pm (z,t) = 0$.

For simulations for the longer resonator, we used the parameters of the As$_2$Se$_3$ chalcogenide fiber resonator that was investigated in reference \cite{Buttner2014} with length ${L_1} = 56{L_B} \approx 38$ cm. The parameters of this resonator are given by: $R = 22.6$ \%, $\alpha=0.84$~dB/m, $n=2.81$, $g_0= 108$ m$^{-1}$W$^{-1}$, $\gamma=1.8$ m$^{-1}$W$^{-1}$,  $\Omega_B/2\pi=7.805$~GHz, $\Gamma_B/2\pi=13.3$~MHz and ${v_a} = 2200$~m/s, which results in $\delta  \approx 0.05$. For the shorter resonator we chose ${L_2} = 6{L_B} \approx 4$ cm. In order to be able to compare the simulation results for the long and the short resonators more easily, we assumed that for the short resonator the interaction area is 10 times smaller compared to the longer resonator, such that ${g_0}$ and $\gamma $  are increased by a factor of 10. We also assumed that the attenuation coefficient $\alpha $ is 10 times larger in the shorter resonator. 

We calculated the strength parameter $\psi$ of the noise terms using the expression given in equation~18 in reference~\cite{Boyd1990} using the following parameters: waveguide temperature $T=293$ K, density of As$_2$Se$_3$ \cite{Abedin2005} $\rho_0=4640$ kg/m$^3$ and interaction area \cite{Buttner2014} $A=56$~$\mu$m$^2$. However, it should be noted that the simulation results discussed in this section are relatively insensitive to the exact noise strength. Due to the presence of optical feedback, the powers of the Stokes waves are rather determined by an interplay of gain and loss.

For our qualitative investigation of the importance of the different FWM interactions with regards to achieving phase-locking, we performed a two dimensional parameter scan of the parameters $\Phi $  and ${P_{\text{in}}}$, and solved the coupled-mode equations for each pair of parameters. We scanned the coupled input power ${P_{\text{in}}}$  from 0.45 W to $0.75$ W in steps of 0.025 W and the phase shift $\Phi $  from  0 to $2\pi$ (which corresponds to tuning the pump frequency $\omega_0$ by one FSR from one cavity resonance to the next one) in steps of $0.05\pi$. The coupled-mode equations were integrated over a time interval of  $t = 4.0\text{ }\mu $s, which is much longer than the phonon and the photon lifetimes in the cavity that determine the time scales of the transient dynamics. 

\begin{figure}[!t]\centering
  \includegraphics[width=0.65\linewidth]{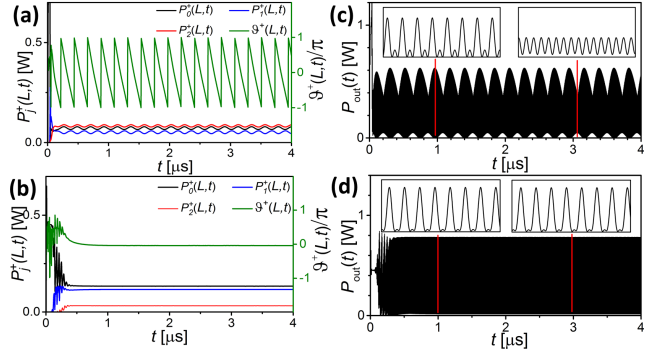}
  \caption{Computed temporal evolution of the powers of pump, first and second-order Stokes waves $P^+_j(L,t)$ and the phase-dispersion $\vartheta^+(L,t)$ at the end of the fiber, for the input power $P_{\text{in}}=0.7$ W, and two different values of $\Phi$: (a) $\Phi=0.52\pi$ (b) $\Phi=1.4\pi$. (c) and (d) show the computed waveforms corresponding to (a) and (b), respectively. }
    \label{fig:theory_temporal}
\end{figure}

Figures~\ref{fig:theory_temporal} (a) and (b) show two typical evolutions of the powers $P_{j}^ + (L,t)$  ($j = 0,1,2$) and the phase-dispersion $\vartheta  ^+ (L,t)$ of the forward propagating waves at the boundary $z = L$ obtained from the simulation. Figures~\ref{fig:theory_temporal} (c) and (d) on the right shows the corresponding time-domain output waveform calculated with 
\begin{align}
P_{\text{out}}(t)=(1-R)\left| {\sum_{j=0,1,2} {A_j^+(L,t)e^{i(k_jL-\omega_jt)} }}\right|^2.
\end{align}
The insets in the plots on the right show 1 ns-long zoomed-in sections at different times. Both simulations were performed for the longer resonator for $P_{\text{in}}=0.7$ W. For the parameter $\Phi$, we chose the values $0.52\pi$ and $1.4\pi$. Figures~\ref{fig:theory_temporal}~(a) and (c) illustrate the case where the system does not reach a phase-locked steady state. Here, the phase-dispersion $\vartheta  ^+ (L,t)$  drifts with time. This also results in an oscillation of the powers  $P_{j}^ + (L,t)$ since the powers of the waves are coupled to the phases via FWM. The obtained output waveform in Fig.~\ref{fig:theory_temporal} (c) also changes with time. Figures~\ref{fig:theory_temporal}~(b) and (d) show the case where the system attained a phase-locked steady state. After an initial power transfer from pump to the first and second order Stokes waves, the powers $P_{j}^ + (L,t)$ and the phase-dispersion $\vartheta  ^+ (L,t)$  reached a constant value, which resulted in a stable waveform shown on the right. 

\begin{figure}[!t]\centering
  \includegraphics[width=0.5\linewidth]{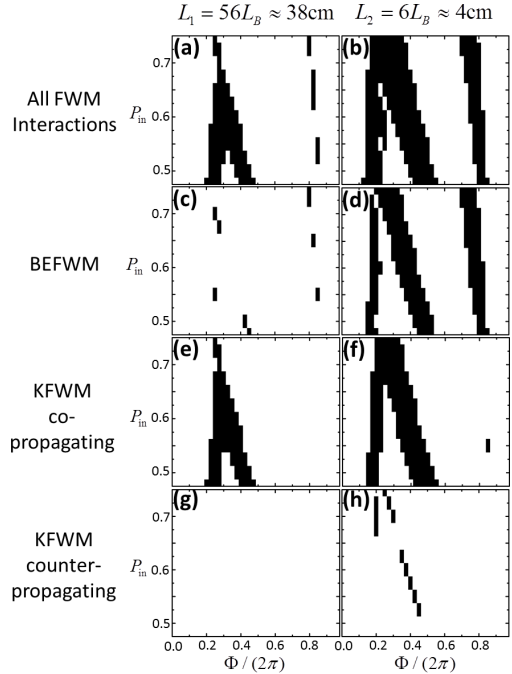}
  \caption{Scan of the parameters $\Phi$  and $P_{\text{in}}$ to identify domains of phase-locking (black). Parameters that lead to phase-locking were determined by integrating the dynamic coupled-mode equations for cascaded BSBS for a sufficiently long time interval and determining whether $\vartheta  ^+ (L,t)$ reached a stable value when the second Stokes wave got sufficiently excited. The left and right column show results for the long and the short cavity, respectively. In order to identify the importance of the different FWM interactions with regards to phase-locking, the parameter scan was performed including all FWM interactions (a), (b); only BEFWM (c), (d); only KFWM of co-propagating waves (e), (f); only KFWM of counter-propagating waves (g), (h). Phase-insensitive BSBS interactions (illustrated in Fig.~\ref{fig:theory_fwm_illustration} (a)), and phase-insensitive terms arising from the Kerr-nonlinearity (self-phase and cross-phase modulation) were kept in the equations for all simulations.
}
    \label{fig:theory_stability}
\end{figure}

The results of the entire parameter scans for the long and the short cavity are shown in Figs.~\ref{fig:theory_stability} (a) and (b). The black area corresponds to parameters for which the dynamic system attained a phase-locked steady state with at least 1~\% of the coupled power in the second Stokes wave $P_2^ + (L,t)$. The simulation results for these parameters are similar to the ones shown in Figs.~\ref{fig:theory_temporal}~(b) and~(d). The white area corresponds to parameters for which the phase-dispersion $\vartheta  ^+ (L,t)$  did not reach a stable value similar to the case shown in Figs.~\ref{fig:theory_temporal}~(a) and (c). The white area also includes parameters for which the second Stokes wave was not efficiently excited. The phase-dispersion  $\vartheta  ^+ (L,t)$ was judged to have reached a stable value when its standard deviation fulfilled  $\sigma (\vartheta  ^+  (L,t)) < {10^{ - 5}}$  over the final $1.5\text{ }\mu $s of the simulation time interval.

To investigate the importance of the different FWM interactions with regard to phase-locking, we performed the same parameter scan for both fiber lengths three additional times. Here, we only included FWM terms corresponding to BEFWM (Figs.~\ref{fig:theory_stability} (c), (d)), only KFWM of co-propagating waves (Figs.~\ref{fig:theory_stability}~(e), (f)) and only KFWM of counter-propagating waves (Figs.~\ref{fig:theory_stability}~(g), (h)), respectively. The terms representing the phase-insensitive BSBS interactions (illustrated in Fig.~\ref{fig:theory_fwm_illustration} (a)), and phase-insensitive terms arising from the Kerr-nonlinearity (self-phase and cross-phase modulation) were kept in the equations for all simulations.

The analysis shows that in the long resonator (Fig.~\ref{fig:theory_stability}, left column), mainly KFWM of co-propagating waves leads to phase-locking whereas in the case of the short resonator (Fig.~\ref{fig:theory_stability}, right column),  KFWM of co-propagating waves and BEFWM can both efficiently phase-lock the waves for a large parameter space. KFWM of counter-propagating waves does not seem relevant for both resonator lengths in the presence of the other FWM interactions. For the long cavity, it did not lead to phase-locking for any parameters. In the short cavity, phase-locking was observed only for a very small parameter space that is included in the parameter space for phase-locking by BEFWM (Fig.~\ref{fig:theory_stability}~(d)).

The results can be understood considering that the coherence length of BEFWM and Kerr-FWM of counter-propagating waves is $L_B$. The long resonator has a length $L_1\gg L_B$. Therefore, the contributions of the non-phase-matched interactions almost completely cancel  along the resonator length. In the case of the shorter resonator, which is only a few times longer that the coherence length $L_B$, the non-phase-matched BEFWM interactions can have a strong contribution (Fig.~\ref{fig:theory_stability}~(d)). This is possible, despite the resonator length being longer than $L_B$, because the optical and acoustic fields are not uniformly distributed along the resonator length due to its low finesse. The amplitudes can vary on a length scale of $L_B$ since the gain was increased in the case of the shorter resonator.

Note that the main reason for the difference of the results shown in Figs.~\ref{fig:theory_stability}~(e) and (f) are the different ratios between phonon and photon lifetime for the long and the short resonators. In this case, no interactions were included that are not phase-matched and the waves experience a comparable gain per transit. However, in the short resonator, the photon lifetime is shorter, which broadens the linewidth of the cavity resonances and leads to a different ratio between cavity linewidth and Brillouin gain linewidth.

\section{Discussion}\label{sec:fiber_discussion}

In this paper, we derived the dynamic coupled-mode equations for cascaded BSBS in fibers including all KFWM and BEFWM interactions. Coupled-mode equations were obtained by expressing the acoustic field with a single acoustic wave and as a superposition of several acoustic waves with different propagation constants. We used the derived equations to numerically  simulate cascaded BSBS in a low-finesse, chalcogenide, FP fiber resonators of two different lengths and qualitatively investigated the relative importance of the different FWM terms with regards to phase-locking of the optical waves. In the longer resonator, corresponding to a previous experiment based on a 38 cm long chalcogenide fiber resonator \cite{Buttner2014}, KFWM of co-propagating waves is most relevant for phase-locking. In contrast, in the shorter resonator of 4~cm length, KFWM and the phase-sensitive coupling of several optical waves via BEFWM contributed to phase-locking.

In the model described in this paper, dispersion has been neglected since we assumed that the interaction takes place over a narrow optical bandwidth as in the experiments described in Refs. \cite{Buttner2014,Buettner2014b}. For numerical modeling of broad BFCs that are generated by the interplay of BSBS and the Kerr-nonlinearity, as demonstrated in reference \cite{Braje2009}, dispersion would have to be considered. 

Equations~\ref{eq:bfc_fiber_aj_with_qls} and~\ref{eq:bfc_fiber_ql} were derived by expressing the electrical and the acoustic field as a superposition of several co-propagating waves with different propagation constants. 
This approach is advantageous for numerical modeling of waveguides that are long enough that KFWM of counter-propagating waves and BEFWM can be neglected, e.g.~long silica fiber resonators. This can be understood considering that simulating the interaction of counter-propagating optical waves requires performing an integration at each spatial grid point of the resonator for each time step. Describing the acoustic field as a single wave requires the spatial grid to be fine enough to resolve the modulations of the acoustic wave on a length scale below $L_B$, in order to include BSBS of higher order Stokes components (see equation~\ref{eq:theory_cme_acoustic_1acoustic}). On the other hand, when the acoustic wave is split in components with different propagation constants, as in equation~\ref{eq:bfc_fiber_ql}, the amplitudes of the acoustic waves $Q_l^\pm$ vary on the same length scale as the amplitudes of the optical waves, which can be tens of meters, e.g.~in the case of a long silica fiber. This allows a significantly coarser spatial grid for the numerical simulation, which can decrease the required computation time by several orders of magnitude.  

In principle, cascaded BSBS can also be described by representing both the electrical and acoustic fields with one forward and one backward propagating wave. The generation of new frequencies then manifests as modulations of the amplitudes of the waves. This could be advantageous for modeling short resonators as it significantly reduces the number of coupling terms since this number is then independent of the number of generated Stokes waves. A spatial discretisation that is fine enough to resolve the amplitude modulations, representing higher order Stokes waves, would be necessary. For BSBS on very short length scales, the propagation of the acoustic wave may also have to be considered if the amplitudes of the acoustic waves vary on lengths scales that are on the order of the decay length of the acoustic wave ($v_a\tau_B\sim 50\text{ }\mu$m).

\section*{Funding Information}

Australian Research Council (ARC) (CE110001018, DE130101033, FL120100029).

\section*{Acknowledgments}

The authors thank Christian Wolff for helpful discussions and Bj\"{o}rn Sturmberg for help with parallelizing the numerical routine.

\end{document}